# 3D IC optimal layout design

# A parallel and distributed topological approach

Katarzyna Grzesiak-Kopeć[a]* and Maciej Ogorzałek[b]

[a,b]*Department of Information Technologies, Jagiellonian University in Kraków, Poland*

*katarzyna.grzesiak-kopec@uj.edu.pl

The task of 3D ICs layout design involves the assembly of millions of components taking into account many different requirements and constraints such as topological, wiring or manufacturability ones. It is a NP-hard problem that requires new non-deterministic and heuristic algorithms. Considering the time complexity, the commonly applied Fiduccia-Mattheyses partitioning algorithm is superior to any other local search method. Nevertheless, it can often miss to reach a quasi-optimal solution in 3D spaces. The presented approach uses an original 3D layout graph partitioning heuristics implemented with use of the extremal optimization method. The goal is to minimize the total wire-length in the chip. In order to improve the time complexity a parallel and distributed Java implementation is applied. Inside one Java Virtual Machine separate optimization algorithms are executed by independent threads. The work may also be shared among different machines by means of The Java Remote Method Invocation system.

**Introduction**

Optimal layout design is one of the main engineering design tasks. Optimal is a *key word* for a design that optimizes a set of goals and satisfies a set of constraints in the same time. Both goals and constraints are often conflicting requirements like miniaturization and usability, low production cost and high quality or practical and aesthetic reasons. *Layout design* may be found in the literature under different headings, e. g. packing, packaging, spatial arrangement, floor-layout, configuration or component layout. Thus, the search space of *optimal layout design* solutions is composed of: (1) design components and their topological connections and (2) design objectives and constraints. Usually, this space is too large to perform an effective deterministic search procedure and some heuristic algorithms are applied to obtain a globally near-optimal solution. Moreover, the ongoing technological development causes the enormous increase of systems complexity and the number of components to consider. The electronic industry is the best example of this progress. In the year 2017, the transistor count (a number of transistors on an integrated circuit (IC)) exceeded 19 billions! (Mujtaba 2017). Hence, the physical arrangement of chip components comprises a myriad of conditions.

Even though the concept of 3-dimensional (3D) circuits integration was first demonstrated as early as in 1979 (Geis et al. 1979) and attracted researchers from industries as well as academics, the 2-dimensional (2D) technology was scaling so well, that there was no market pull to develop it. Nowadays 3D ICs design is being reconsidered and become a sine qua non for the silicon world. It definitely improves circuit blocks packing density and dramatically decreases the total interconnect wire length. Along with the interconnect wire length reduction the power consumption is decreased as well. The third dimension allows heterogeneous technology integration such as digital and analog. Different components can be manufactured separately according to their technology and then stack together on a single chip. And last but not least, 3D design not only gives a smaller footprint but the total volume minimization, which is very suitable for commonly used mobile devices (Dong and Xie 2009). There are two major groups of 3D integration technologies: integration using chip stacking and Through Silicon Vias (TSVs), and native 3D integration. The latter approach is still in its infancy. Despite the abundance of electronic design automation tools for 2D integration, there is still a great need for the specific 3D tools, methods and flows to support the growth of the 3D IC market. Most available software packages are extensions of those used for planar (2D) design (De Micheli et al. 2011).

The proposed native 3D layout design approach introduces three separate design representation layers, namely the semantic layer, the presentation layer and the optimization control one (Grzesiak-Kopeć and Ogorzałek 2014). Possible solutions are generated with use of a simple shape grammar supervised by an intelligent derivation controller. The shape grammar is defined by the designer, who also provides a specific design knowledge in a form of predicates. The predicates are fed into the generation and optimization procedures. The total wire-length of a generated result may be further optimized adopting a knowledge intensive 3D ICs layout hypergraph representation described in (Grzesiak-Kopeć and Ogorzałek 2015a), together with the elaborated neighbourhood optimization heuristics presented in (Grzesiak-Kopeć and Ogorzałek 2015b).

This article mainly deals with the total wire-length minimization. The main novel contribution is the *volume optimization* procedure for eliminating gaps/empty spaces in the generated 3D structure. The 3-step intelligent wire-length optimization approach is illustrated by the example of application to the MCNC benchmark circuits (MCNC) using a parallel and distributed Java implementation. First, the knowledge intensive 3D ICs layout hypergraph representation together with the elaborated neighbourhood optimization heuristics are introduced. Then, the wire-length extremal optimization is described. After that, the procedure of the volume optimization together with the parallel and distributed implementation are explained. Finally, the proposed wire-length optimization heuristics is applied to the MCNC set of benchmark circuits and the experimental results are reported.

**Related work**

The physical arrangement of components plays a crucial role in integrated circuit design. It directly affects circuit performance, area, reliability, power, and manufacturing yield (Kahng et al. 2011). It enables assessment of system architecture decisions and estimation of delay and congestion caused by wiring (Wang et al. 2009). The today's 3D ICs technology still has various limitations such as a layer-like structure, where a number of device layers is restricted and the inter-layer height is fixed. Still, a quasi-3D placement problem is much more complex than a true-2D one. Generally, the placement problem is known to be NP-hard (Garey and Johnson 1979, Lengauer 1990). Adding an extra dimension to the solution space definitely increases the difficulty of the circuit design task in many aspects. When taking into account a placement problem with *n* components and *k* layers, a 2D solution may be divided into $\frac{n^{k-1}}{(k-1)!}$ different *k*-layer 3D floorplans (Li et al. 2006). Thus the solution space complexity raises by this number times and results in longer searching time and/or worse placement quality. Common techniques for global placements are: partitioning-based algorithms, analytic techniques and stochastic algorithms (Kahng et al. 2011, Hentschke 2007).

Recursive partitioning are constructive techniques with average CPU requirements and versatility. The netlist and the layout are recursively divided into smaller sets/problems according to a cut-based cost function until the parts are small enough to be solved optimally (Taghavi 2005). Common algorithms used to minimize the number of cut nets are the Kernighan-Lin (Kernighan and Lin 1970) and the Fiduccia-Mattheyses algorithm (Fiduccia and Mattheyses 1982). The most popular is the latter which is both computationally effective (a linear time heuristics) and easily adjustable to different fitness functions. It consists of three stages: coarsening, initial partitioning and uncoarsening. During the coarsening phase, the netlist is successively contracted until it is small enough to be plausibly partitioned in the initial partitioning phase. The selected strategy strongly determines general quality of the partition. The contraction is reversed at the time of the uncoarsening phase. The achieved initial partition is mapped to the more comprehensive netlist graph and the solution is improved by a local search algorithm.

Analytic techniques, such as quadratic placement and force-directed placement, are constructive ones with relatively low CPU requirements and average versatility (Eisenmann and Johannes 1998, Obermeier and Johannes 2004). They use an objective quadratic or otherwise non-convex function, that can be minimized/maximized via mathematical analysis. Quadratic placement is a two-stage approach. The first stage is a global placement that minimizes the quadratic function with respect to the component centers. The obtained overlapping (illegal) solution is corrected during detailed placement to give a final placement result. A special case of quadratic placement is the force-directed placement where the mechanical mass-spring system analogy is used to represent components and wires. The attraction force between components is directly proportional to their distance. The goal is to reach a placement in a state of force equilibrium.

Stochastic algorithms introduce a random factor into the cost function optimization procedure. The best known stochastic placement algorithm is simulated annealing (SA). SA is an iterative approach with high CPU requirements where not much memory but a long execution time is needed to reach the desired solution (Sechen 1988, Wong et al. 1988, Taghavi et al. 2005, Chen and Chang 2006). A generated initial placement is perturbed until the annealing process reaches an equilibrium state or the algorithm stops after a prescribed number of iterations.

Today's global placement algorithms model wire length with mathematical functions and use numerical methods to optimize them (Kahng et al. 2011). The components actual dimensions are initially ignored in order to find a seed placement. After that, they are gradually introduced into the optimization procedure to prevent unbalanced densities and routing congestion. The most popular approaches are based on analytic techniques and nonlinear optimization. The constantly growing complexity of microelectronic systems implies growing importance of partitioning. That is why wire length optimization algorithms are often combined with netlist clustering. The system is divided into some critical parts to speed up and improve the design cycle. Some of the modern placers are: APlace (Kahng et al. 2007), Capo (Roy at al. 2005), FastPlace 3.0 (Viswanathan et al. 2007), mFAR (Hu et al. 2005), mPL6 (Chan et al. 2006), and simPL (Kim et al. 2012).

**Intelligent wire-length optimization**

Aside from 3D technological hurdles, the interconnect wire-length minimization is one of the crucial circuit design requirements. We propose a 3-step optimization approach (Figure 1) where the only input of the task is a netlist given in a YAL file format (MCNC). A netlist description defines the connectivity of an electronic circuit together with the parameters (like dimensions and pins) of the devices. In the first step, a YAL file is parsed into a layout hypergraph, which is a knowledge intensive representation of a netlist structure and its components (Grzesiak-Kopeć and Ogorzałek 2015a). In the second step, a topological partitioning procedure using the extremal optimization is applied to the layout hypergraph in order to find a special arrangement of components. Finally, in the third step, a parallel distributed volume optimization is performed by squeezing the intermediate layout solution. The relative positions of the components are preserved while the gaps and empty spaces are removed.

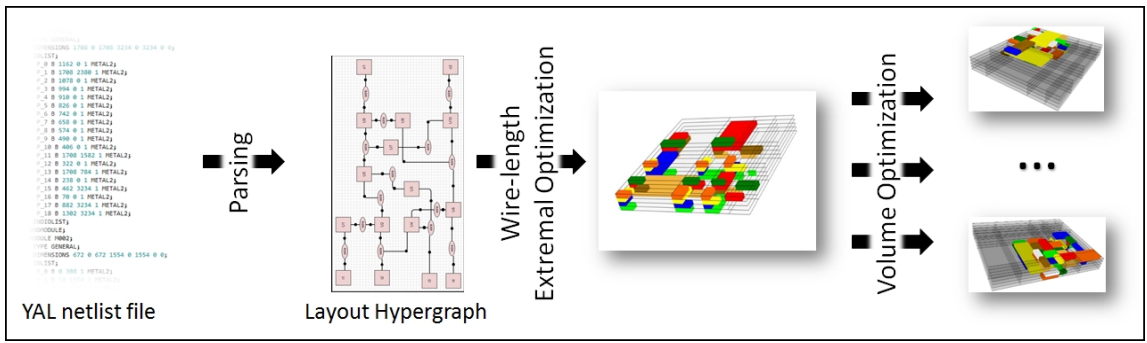

Figure 1. The 3-step intelligent wire-length optimization approach.

*Transforming netlist into a layout hypergraph*

In electronic design, a netlist is a text description of the electronic circuit connectivity. It consists of a collection of several related lists: a list of terminals (pins), a list of instances (components) and a list of signals connected to terminals (connections). It may also contain some attribute information. The connectivity information from a netlist may be formally represented in the form of a graph with appropriate semantic mappings (Mony et al. 2004).

Graphs are data structures especially useful to represent different relational issues in a variety of systems (such as electric circuits, traffic, chemical processes or social networks). In the basic definition a graph depict only binary relations but its generalization, called a hypergraph, effectively abolish this limitation. Recently hypergraphs gain a lot of attention because of their applicability to Web information systems, social networks, document centred information processing and many others service-oriented systems (Molnár 2014). Since the connectivity/wires of the electronic circuit, described by a netlist, can be straightforward mapped into a hypergraph (Figure 2 A, B), they are also used for ICs layout design (Karypis et al. 1999).

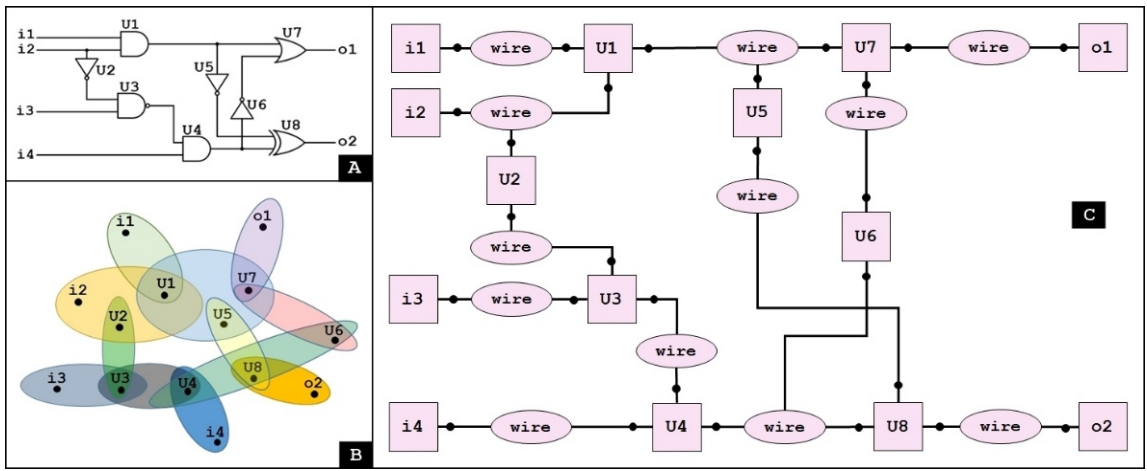

Figure 2 The example: (A) a simple logic circuit, (B) the corresponding hypergraph, (C) the corresponding layout hypergraph.

In our approach, the layout hypergraphs introduced in (Grabska et al. 2006) to depict floor-layouts in architectural design, are used. A layout hypergraph consists of a finite set of vertices (nodes), a finite set of hyperedges and labelling functions that allow assignment of attributes to either nodes or hyperedges. Nodes do not represent any entities but some pin points that are used for fixing represented components together. Both components and relations are depicted by dedicated hyperedges, namely component hyperedges and relation hyperedges. In Figure 2C rectangles represent components hypergedes and ellipses illustrate wire relation ones. Component hyperedges are labelled by the corresponding component names (i1 – i4, U1 – U8, o1, o2). There are nine binary wire relations and three 3-tuples. Black dots denote nodes that are pin points (terminals). If required, a hierarchy relation may be introduced with a use of a child nesting function. The formal definition of such a layout hypergraph can be found in (Ślusarczyk 2012, Grzesiak-Kopeć et al. 2017).

The MCNC block packing instances given in a YAL file format is the most common set of benchmark circuits for floorplanning and placement problems (MCNC). Parsing a YAL file and building a corresponding layout hypergraph is pretty straightforward. The component descriptions are given in appropriate MODULE sections together with their dimensions (DIMENSIONS) and possible terminals (IOLIST). The circuit connectivity is defined in the NETWORK section. Each line specifies a named instance of a module (component) that is a part of the circuit, together with its signal bindings. A single signal identified by its name is represented by a single wire hyperedge in a layout hypergraph. As the number of nets in considered examples is very large (see Table 1), the layout hypergraph drawing of such a circuit would be completely unreadable and therefore is omitted.

*Wire-length Extremal Optimization*

In the context of IC design, a hypergraph representation is applied to solve the wire-length minimization task. The commonly implemented recursive partitioning algorithm splits netlists into parts, just like in the case of 2D floorplans, using only a single dimension (Ababei et al. 2005). A hypergraph is divided into $k$ disjoint nonempty partitions in such a way that the total vertex weight in each partition is balanced and the sum of hyperedge weights that are cut between partitions is minimized. The most popular and the most computationally effective (a linear time heuristic) graph partitioning algorithm is Fiduccia-Mattheyses one (Fiduccia and Mattheyses 1982). It starts with the coarsening phase during which the graph is progressively reduced to be small enough to likely undergo optimal partitioning in the initial partitioning phase. The last stage is the uncoarsening phase when the reduction is reversed. The final result of this approach is highly determined by the applied coarsening strategy. Better quality of the partitioning is achieved when the smaller graphs exhibit similar structural properties to the propertied of the bigger ones.

Unfortunately, the approach with a dimension limiting to the $k$-way partitioning often fails to find a near optimal solution in 3D. Hence, the 3D topological partitioning of the layout hypergraph has been proposed (Grzesiak-Kopeć and Ogorzałek 2015b). Instead of striving for the minimal and balanced cut of the graph, a topology-oriented neighborhood grouping is performed. Bearing in mind the grid-like chip topology, the diamond-shaped von Neumann neighborhood is considered. The Manhattan distance is used to measure the distance between two cells in the IC grid. The shorter adjacent component wire connections are achieved, the better total wire-length solution is obtained. Adopting the building block hypothesis, where the optimal global solution is the sum of the optimal local neighbourhoods, entails the selection of the Extremal Optimization (EO) implementation (Boettcher 2000).

The EO is a co-evolution approach for optimization problems inspired by self-organized critical (SOC) models and the Bak-Sneppen co-evolution model (Bak and Sneppen 1993). The distinguishing trait of this approach is evolving only a single solution $S=\{x_1, x_2, ..., x_m\}$ instead of the whole population of possible solutions. Every solution feature ($x_i$) has its own fitness value and the sum of these values gives the quality of the entire solution $S$. In each step, the evolution procedure eliminates the least desirable feature ($x_i$) of $S$ by generating a new random solution $S'$ where ($x_i$) is altered. Such a purely random strategy may cause a deadlock in some implementations. To overcome this disadvantage, a $\tau$ control parameter may be introduced that enables a ranking selection instead of a simple random change of the worst adopted features.

During the $\tau$-Extremal Optimization ($\tau$-EO) all the features are sorted in ascending order according to their fitness values. Step by step, a new solution $S'$ is achieved by swapping two features from a probability distribution: $P(k) \propto k^{-\tau}$, $1 \leq k \leq m$. In such a $\tau$-controlled approach, almost all features are significantly better in the self-organized critical state than initial ones. Their fitness values are highly correlated to the quality of their neighbours. The generation procedure maintains good values if their neighbouring features are not low-adapted. The adaptation process improves the solution quality applying a hill climbing technique for some time to crash it suddenly (avalanche) as described by punctuated equilibrium. The avalanches act as the mutation operator in other evolutionary approaches and shift the search process to escape the local optima. They are emergent behaviours of the negative feature selection that provide diversity and assure large variation at any stage of the optimization procedure (Boettcher 2000).

In the case of the IC total wire-length optimization task, the search space is limited by a predefined maximal chip volume X×Y×Z. Neglecting original dimensions, the chip components are placed in this cuboid in such a way that each cell $(x,y,z) \in X \times Y \times Z$ is either occupied by a component or empty. Empty cells that are not located in the boundary of the chip are not welcome and, if possible, should be eliminated from the final solution. Components designate the solution features and they are evaluated on the basis of the range of their neighbourhoods. The optimal fitness of each component is the minimal range neighbourhood containing all of its neighbors (for details see (Grzesiak-Kopeć et al. 2015). In this way, the optimal fitness of the whole solution $S$ equals the number of components.

The τ-EO results for the MCNC benchmark are presented in (Grzesiak-Kopeć et al. 2017). Generated layout examples of the MCNC instances (*apte*, *xerox*, *hp*, *ami33* and *ami49* ) are depicted in Figure 3. The cell sizes are adjusted to the component dimensions and numerous unwanted gaps are visible.

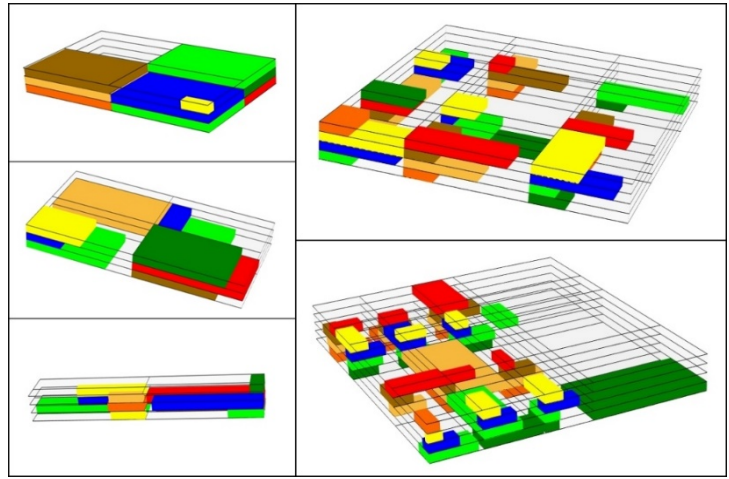

Figure 3 The τ-EO results for the MCNC benchmark.

*Volume Optimization*

The τ-EO optimal topological partitioning of the layout hypergraphs generates solutions in a predefined 3D grid cuboid and neglects the actual dimensions of the modules. In such an approach some cells in the grid usually remain empty. In most cases it is possible to eliminate at least some of these gaps without violating the optimal spatial arrangement of components. In other words, the relative arrangement of components is preserved.

*Squeezing*

The elaborated step-by-step squeezing procedure moves the chip components towards a predefined rallying point *P* along either the X or the Y axis. The *Single Component Move Algorithm* proceeds in the following way. If a component cannot be moved in a selected direction it is added to *immobileComponentsX* or *immobileComponentsY*, respectively. A component that is immobile in both directions is blocked and removed from *possibleMovesQueue*. A rallying point *P* is selected arbitrary by the designer. It can be a center of the chip or one of the corners of the chip bounding box, or a center of mass of the chip or any other point in 3D. A current move of a component is selected randomly from a queue of possible moves (*possibleMovesQueue*). If a new component position is not closer to *P* it will be rejected. Otherwise, it is verified whether a new position collides with other components. When the currently sliding component collides with any component that is blocked (immobile), its moves in the selected direction are also blocked. Summing up, a component is translated if and only if it is approaching the point *P* and no collision with other components is recognized. The algorithm pseudocode is presented in Figure 4.

```
Single Component Move Algorithm
                    P:   a predefined rallying point
 possibleMovesQueue :   all the possible moves of the components
 immobileComponentsX:   all the components that moves in X direction are
                        unwanted
 immobileComponentsY:   all the components that moves in Y direction are
                        unwanted
 immobileCommponents = immobileComponentsX ∩ immobileComponentsY
 movePossible = true
```
```
BlockComponent( component ){
  - indicate the direction component was moved towards
  - add component to immobileComponentsX or immobileComponentsY accordingly
  - if component is in immobileComponentsX and immobileComponentsY then
    - add component to immobileComponents
  - remove component and direction from possibleMovesQueue
}
```
```
Until movePossible repeat:
  (1) Shuffle possibleMovesQueue
  (2) currentMovesQueue = possibleMovesQueue
  (3) movePossible = false
  (4) Untill !currentMovesQueue.isEmpty repeat:
    (a) get currentComponent and move direction from currentMovesQueue
    (b) set new position for currentComponent
    (c) if new position is not closer to P then
        - BlockComponent(currentComponent)
        - continue
    (d) if new position collide with other components then
        - indicate the components that collide with currentComponent
        - if any component from immobileComponents is indicated then
          - BlockComponent(currentComponent)
    (e) else
        - set new position for currentComponent
        - put currentComponent in front of currentQueue according
          to probability P1
        - put currentComponent in rear of currentQueue according
          to probability P2
        - movePossible = true
```

Figure 4 The Single Component Move Algorithm pseudocode.

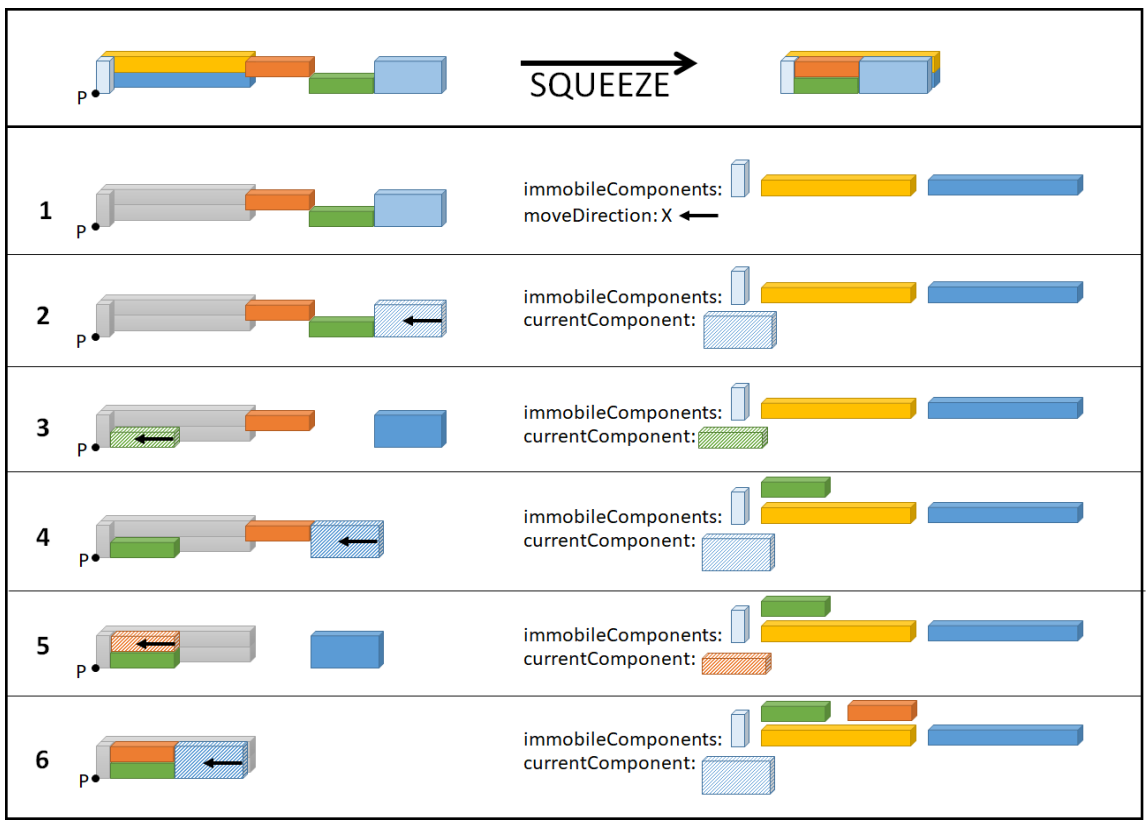

Figure 5 The example arrangement of components before and after the squeezing; (1)-(6) selected intermediate steps.

Let us consider the squeezing example in Figure 5. A rallying point *P* is a left-bottom-front corner of the chip bounding box. The first step (Figure 5.1) presents some intermediate situation where three components have already been recognized as blocked (*immobileComponents*). In the following steps, only moves towards the point *P* along the X axis are considered. Unfortunately, the *currentComponent* selected in the next step (the blue one) (Figure 5.2) cannot be moved in the preferred direction due to the collision with another component (green). Since the colliding green component is still marked as mobile, the action is just neglected at this moment and may be reconsidered in the future. In Figure 5.3 the move of the *currentComponent* (green) is admissible and performed. After that, the green component cannot be translated any further and if it is selected in any of the following steps it will be blocked and marked as

*immobileComponent*. In Figure 5.4, the blue component is selected once again and moved towards *P* as close as it is possible – till it collides with the orange one. In the last two situations (Figure 5.5-6), the orange component and the blue one, respectively, are moved to their final locations.

It is also possible to move at the same time a whole bundle of components. Such a move is allowed if: (1) all the components in a bundle can move in the same direction and the move place them closer to *P*, (2) no collision occurs. Applying a bundle approach to the example in Figure 5, the whole squeezing could be performed in two steps instead of five. In the first move, a bundle of three components (orange, green and blue) would be translated towards *P* along the X axis and the orange component would reach its final position. In the second move, a bundle of two last components (green and blue) would be translated to its terminal location.

*Parallel and distributed computations*

Taking into account the grid-like structure of the plausible results, the number of possible moves in each step is upper bounded by the number of components multiplied by 2 (the X and the Y direction). The final result is dependent on the actual sequence of moves because they are strongly correlated with each other. Hence, the randomize selection of the next action has been introduced and the *possibleMovesQueue* is shuffled before the current move selection. Furthermore, in order to maximize the number of analysed available solutions, parallel and distributed computations have been applied.

Distributed computations are shared among autonomous computers that communicate with each other in order to achieve a common goal. The computers are independent which means that they do not physically share processors or memory. They communicate and coordinate their work using messages passed over a network. They may play different roles and be organized in different ways. There are two predominant architectures: client-server and peer-to-peer architecture (Coulouris et al. 2011).

The volume optimization application is written in Java. Multiple optimization procedures are executed in parallel either by independent threads inside one Java Virtual Machine or are shared between many machines by means of The Java Remote Method Invocation system. In the latter case, the client-server architecture is adopted. One instance of the program plays the role of the server and the others are run in the client mode. The server generates and spools optimization tasks, while the clients fetch and do the jobs (Figure 6). The client list is managed dynamically and at any time machines are able to join or leave the computing system.

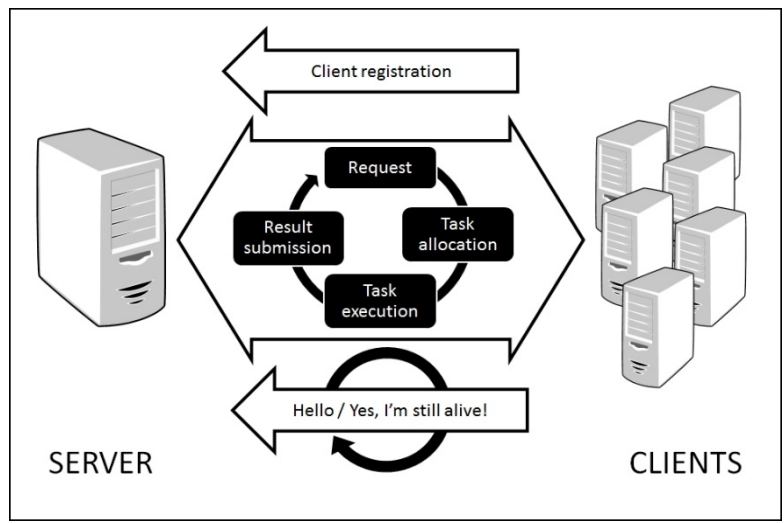

Figure 6 The parallel and distributed calculation scheme of the squeezing.

The proposed system architecture is fault tolerant. All the tasks are unified and kept in a single queue. Each type of the task is elaborated by a dedicated plugin that is matched by the object type. The server distributes the tasks among clients and waits for "hello" messages which confirm that the clients are still working. If no "hello" is received from a client for a specified period of time, the client timeout is recorded and its unfinished task is going back to the queue (Figure 7).

The squeezed τ-EO results for the MCNC benchmark are presented in Figure 8. The cell sizes are no longer adjusted to the maximal component dimensions, like in Figure 3. Many unwanted empty spaces are successfully removed while preserving the relative spatial relations of components.

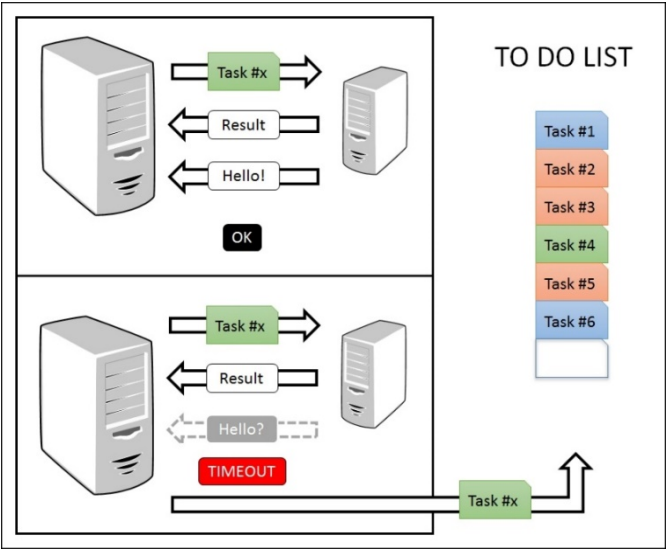

Figure 7 The fault tolerant task realization scheme.

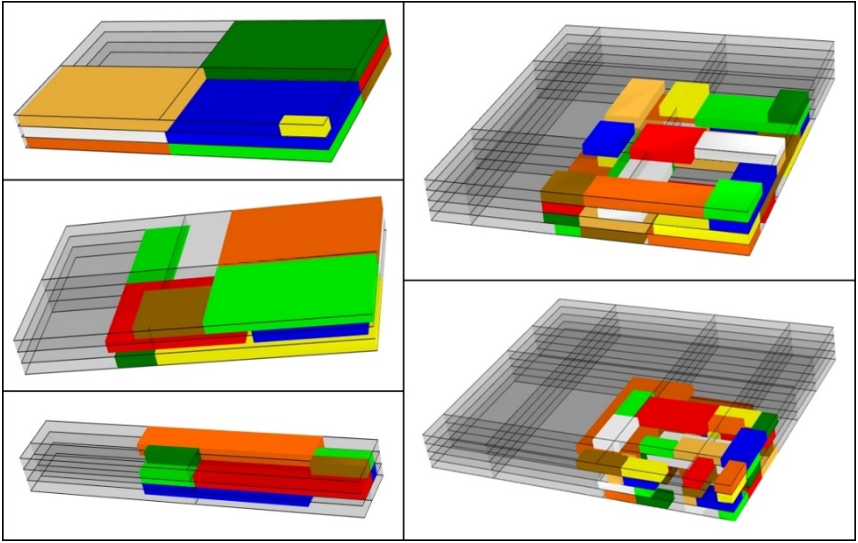

Figure 8 The squeezed τ-EO results for the MCNC benchmark.

**Experimental results**

The MCNC benchmark netlists (MCNC) are one of the most frequently used for floorplanning and placement problems. Five block packing instances are given in a YAL file format. Their characteristics are listed in Table 1, where the columns denote: the file name (YAL datafile), the number of components (Blocks), the number of nets (Nets), the minimal, the maximal and the average number of interconnected components respectively (Neighbors no/ min/ max/ avg). As stated before, by *neighbours* we denote components that comprise a single net. Each component may be a part of many different nets.

Table 1 Characteristics of the MCNC benchmark instances.

| YAL datafile | Blocks | Nets | Neighbors no | | |
|---|---|---|---|---|---|
| | | | min | max | avg |
| apte.yal | 9 | 97 | 8 | 8 | 8 |
| xerox.yal | 10 | 203 | 9 | 9 | 9 |
| hp.yal | 11 | 83 | 5 | 10 | 7 |
| ami33.yal | 33 | 123 | 32 | 32 | 32 |
| ami49.yal | 49 | 408 | 2 | 35 | 18 |

In (Funke et al. 2016) the optimal wire-length in 2D for the three smallest MCNC instances (*apte*, *xerox* and *hp*) was calculated. It also was stated, that computing wire-length optimal packings for the two remaining instances (*ami33* and *ami49*) is still far beyond the realms of possibility. The basic half-perimeter model (HPWL) for a wire-length calculation was applied, where the wire-length of a net is a half of the perimeter of the bounding rectangle that encloses all the pins of the net. It is one of the most widely used approximation schemes. In such a way calculated optimal results together with the results reported in (Funke et al. 2016) for *ami33* and *ami49* are given in Table 2.

Table 2 Optimal wire-lengths for *apte*, *xerox*, and *hp* for the original die size (after (Funke et al. 2012)). Wire-length for *ami33* and *ami49* after (Funke et al. 2016). Values are given in µm.

| YAL datafile | original size | wire-length *optimal |
|---|---|---|
| apte.yal | 10 500 × 10 500 | 513 061* |
| xerox.yal | 5 831 × 6 412 | 370 993* |
| hp.yal | 4 928 × 4 200 | 153 328* |
| ami33.yal | 2 058 × 1 463 | 58 627 |
| ami49.yal | 7 672 × 7 840 | 640 509 |

The extremal optimization procedure was successfully evaluated in (Grzesiak-Kopeć et al. 2015) and different approaches were examined (Grzesiak-Kopeć et al. 2017). The numerical data proved that the initial component layout basically does not matter for the final result as was expected for a fine defined extremal optimization task. In this article, the total wire-length approximation results, calculated after the squeezing volume optimization, are presented. In order to compare the proposed solution with the optimal results in 2D (see Table 2), the HPWL wire-length model approximation was applied. The third dimension was introduced into a formula as the height of the bounding cube of the components which belong to a net. Thus, the wire-length *w(n)* of a net *n* is calculated as follows:

$$w(n) = max_{c',c'' \in n}|c'_x - c''_x| + max_{c',c'' \in n}|c'_y - c''_y| + max_{c',c'' \in n}|c'_z - c''_z| \quad (1)$$

where *c'* and *c''* are components that belong to a network *n* and $c_x$, $c_y$, $c_z$ denote *(x,y,z)* position of a component *c*, respectively. The experimental results are listed in Table 3.

Although many floorplanning approaches are applied to the MCNC set of benchmarks there are only a few that contain comparable wire-lengths in 2D, like (Funke et al. 2016, Liu and Nannarelli 2008). It is caused by the fact that the authors do not use the original die sizes (Table 2) but modify them by scaling, rotating or splitting block modules in order to reduce whitespaces (Nain and Chrzanowska-Jeske 2011, Xie and Zhao 2015). In this way, the chip components are changed and an essentially different layout task is solved. Still, it was reported in (Das et al. 2004), that depending on the number of chip layers, the average 28% to 51% reduction in the total wire-length may be achieved in 3D. Our results presented in Table 4 confirm this premise. Apart from the *ami49* instance, the total wire-length was reduced by 21%-73% compared to the best results in 2D (Funke et al. 2016). Only in the case of *ami49* circuit, the total wire-length is by 10% longer. Yet, it is by 34% better that the 2D result presented in (Liu and Nannarelli 2008).

Table 3 The volume optimization results for the MCNC benchmark original die size. Volumes are given in μm (die height equals 1).

| YAL datafile | τ-EO grid size | volume | wire-length |
|---|---|---|---|
| apte.yal | 2×2×3 | 5 018×4 972×3 | 137 325 |
| xerox.yal | 2×2×3 | 3 864×3 829×3 | 290 183 |
| hp.yal | 2×2×3 | 3 758×3 542×3 | 105 848 |
| ami33.yal | 3×3×4 | 911×1 163×4 | 42 183 |
| ami49.yal | 4×4×4 | 5 769×5 979×4 | 704 135 |

Table 4 The wire-lengths results in μm for the MCNC benchmark instances.

| Article | apte | Xerox | hp | ami33 | ami49 |
|---|---|---|---|---|---|
| This (3D) | 137 325 | 290 183 | 105 848 | 42 183 | 704 135 |
| (Funke et al. 2016) (2D) | 513 061 | 370 993 | 153 328 | 58 627 | 640 509 |
| (Liu and Nannarelli 2008) (2D) | 614 602 | 404 278 | 253 366 | 96 205 | 1 070 010 |
| (Nain and Chrzanowska-Jeske 2011) (3D) | - | - | - | 22 500 | 446 800 |
| (Xie and Zhao 2015) (3D) | - | 297 440 | 124 819 | 27 911 | 547 491 |

In (Xie and Zhao 2015) various perturbations that change the original modules are allowed to handle 3D floorplans, namely rotation and resize. Even though, our results for *xerox* and *hp* are a bit better. However, when bigger benchmarks are considered (*ami33* and *ami49*) our wire-length are by 33% and 22% longer. Taking into account 3D solutions for *ami33* and *ami49* in (Nain and Chrzanowska-Jeske 2011) where some modules are split and their parts are assigned to different device layers, our results are by 47% and 37% worst.

At the first glance, the proposed approach seems to be inferior to others when applied to bigger benchmarks. However, it must be stressed out that changing the original modules changes the whole layout task. Hence, the achieved solutions are incomparable. Our approach is general and does not use any specific knowledge about circuit building blocks except of the netlist. It does not modify the building blocks structure. Considering the generated four-layer 3D floorplan of *ami49* (Figure 8), some may criticize it for too many gaps (white spaces) comparing to the floorplans presented in (Nain and Chrzanowska-Jeske 2011). Therefore, one may assume that when perturbations of original modules are allowed the total wire-length, which is satisfactory right now, will even improve.

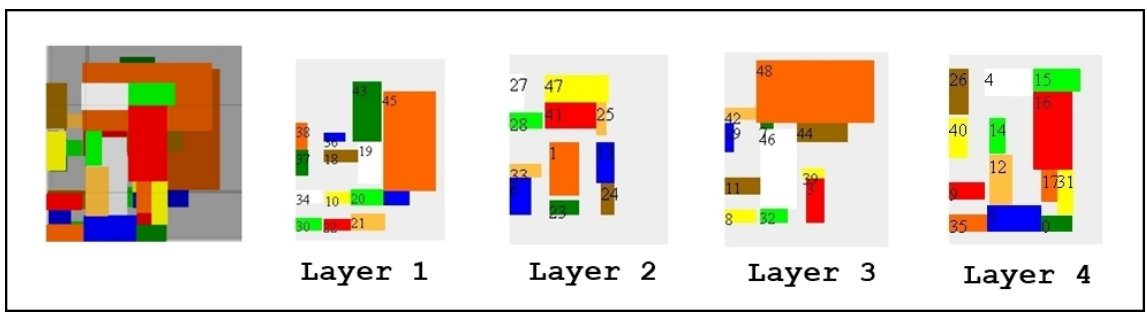

Figure 9 The 3D floorplan of *ami49*.

**Conclusions**

This article presents an original and general 3-step intelligent approach to the total wire-length minimization in the integrated circuits design. In the first step, a netlist YAL file is parsed into the elaborated layout hypergraph representation. After that, a topological partitioning with a use of the extremal optimization is applied to optimize the relative positions of the components in the chip. And finally, a parallel distributed volume optimization is performed. The squeezing procedure is executed in order to minimize the total wire-length of the final solution.

There is no a priori knowledge needed to solve the floorplan puzzle. Only the original block modules sizes and the netlist connectivity information are considered. That is why, the reported numerical results for both the chip volume size and the total wire-length are very promising and encourage to continue this research. In the future work, the knowledge about the circuit building blocks may be incorporated to allow components perturbations. Furthermore, the exact pin points positions used in the wire-length calculation formula would give precise instead of the rough results.